\documentclass[aps,preprint,pre,showpacs,eqsecnum,amssymb]{revtex4-1}
\usepackage{graphicx}
\begin{document}
\title{Energetics and efficiency of a molecular motor model}
\author{Hans C. Fogedby}
\email{fogedby@phys.au.dk}
\affiliation{Department of Physics and Astronomy, University of
Aarhus\\Ny Munkegade 120, DK-8000, Aarhus C, Denmark\\}
\affiliation{Niels Bohr Institute\\
Blegdamsvej 17, 2100, Copenhagen {\O}, Denmark}
\author{Axel Svane}
\email{svane@phys.au.dk} \affiliation{Department of Physics and
Astronomy, University of Aarhus\\Ny Munkegade 120, DK-8000, Aarhus C,
Denmark}

\begin{abstract}

The energetics and efficiency of a linear molecular motor model
proposed by Mogilner et al. (Phys. Lett. {\bf 237}, 297 (1998))
is analyzed from an analytical point of view. The model which is 
based on protein friction with a track is described by coupled 
Langevin equations for the motion in combination with 
coupled master equations for the ATP hydrolysis. 
Here the energetics and efficiency of the motor is addressed
using a many body scheme with focus on the efficiency
at maximum power (EMP). It is found that the EMP is reduced
from about 10\% in a heuristic description of the motor to about
1 per mille when incorporating the full motor dynamics,
owing to the strong dissipation associated with the motor action.
\end{abstract}
\pacs{05.70.Ln, 05.40.-a, 87.16.Nn}.

\maketitle

\section{\label{intro} Introduction}
There is a current interest in the energetics and efficiency of molecular motors
and nano engines. Molecular motors are energy-consuming, non-equilibrium
nanoscale engines encountered in various dynamical processes on the intra- and
intercellular level \cite{Howard97,Alberts94}; for a recent review
of the more physical aspects see, for instance Ref.~\cite{Reimann01}. 
Linear motor proteins like myosin or kinesin are driven by the hydrolysis
of ATP into ADP  and  move along linear polar  tracks like actin filaments or microtubules. 
The motors typically work in an isothermal environment at ambient temperatures
subject to viscous forces.

Modern experimental techniques in biology and biophysics, in
particular single biomolecule manipulation by for example optical
tweezers or micro-needles, and single particle tracking methods,
have yielded considerable insight into the mechanism and the
relevant physical scales in molecular motor behavior, see e.g. 
\cite{Wang98}. The typical size of a molecular motor is
of order 10 - 20 nm, moving with a step size of order 8 nm, e.g.,
kinesin on microtubules, with one ATP molecule hydrolyzed on the
average per step. The velocities of molecular motors range from
nm/sec to $\mu$m/s and the maximum load is of the order of several
pN. The time scale of the chemical cycle is a few ms and the average
energy input from the ATP-ADP cycle of order 15-20 kT. Here kT is the energy
scale of thermal motion at temperature T.

A molecular  motor constitutes  an interesting non-equilibrium system 
operating in the classical regime and is thus directly amenable 
to analysis using methods in statistical physics. Physical modeling of
molecular motors has thus been studied intensively in recent years,
both from the point of view of the fundamental underlying physical
principles and with regard to the specific modelling of concrete
motors \cite{Fisher99,Leibler93,Leibler94,Duke96,Magnasco93,Julicher97,
Astumian97,Astumian99,Astumian02,Ambaye99,Ajdari94,Norden02}. 
The most common statistical approach to molecular motors is 
the ratchet model \cite{Julicher97,Reimann01} modelling the
periodically alternating energy landscape  felt by the motor
during its cycle. An alternative motor model can be based on protein friction
where the motor during its cycle is in contact with a track.
\cite{Tawada91,Leibler93,Jannink96,Brokaw97,Julicher97,Imafuku96}.

In recent work Mogilner et al. \cite{Mogilner98} have studied a specific
protein friction model. The motor is represented by two
coupled over damped  oscillators driven by a two-step Markov process
alternating between a relaxed and a strained state of the
oscillators. The motor is embedded in a thermal environment represented by
additive white noise. The  subprocesses are associated with
internal conformational changes of the motor protein. One
subprocess is slow, allowing protein friction to act,
while the other is fast and only subject to solvent friction. By
means of a numerical analysis Mogilner et al. show that the system
acts like a motor and can carry a load. However, unlike the
ratchet models, which operate with an attachment to a periodic
polar protein filament, the model of Mogilner et al. only needs a
`passive' groove in order to perform directed motion. The
motion comes about due to the asymmetric internal
velocity fluctuations which are then rectified by protein
friction. In that sense, it is a robotic model of molecular
motors.

Recently, there has been an interest in the efficiency of
molecular motors and nano engines \cite{Wang08,Parmeggiani99,
Magnasco93,Golubeva12a,Golubeva12b,Schmiedl08a,Schmiedl08b,vdenbroeck12,Tu13}. 
Unlike heat engines where the
efficiency is bounded by Carnot's law, see e.g. \cite{Reichl98},
the molecular motors work in an isothermal environment at ambient
temperatures and the efficiency can in principle reach unity. The
efficiency of a motor is given by
\begin{eqnarray}
\eta=\frac{p}{\epsilon_{\text{in}}}~, \label{eff}
\end{eqnarray}
where $p$ is the output power, i.e., the work done per unit time on
the surroundings. Likewise, $\epsilon_{\text{in}}$ is the input energy
rate. The conservation of energy implies
\begin{eqnarray}
\epsilon_{\text{in}}=p+q_{\text{out}}, \label{econ}
\end{eqnarray}
where $q_{\text{out}}$ is the dissipation rate. Here $\epsilon_{\text{in}}$ refers
to the energy input to the motor arising from the ATP hydrolysis, whereas
$q_{\text{out}}$, the dissipation rate,  refers to the frictional effects during the
motor operation. $\epsilon_{\text{in}}$ and $q_{\text{out}}$ refer to two different energy
reservoirs and there is thus no conflict with the second law of thermodynamics.
The theoretical upper limit $\eta=1$ requires the absence of irreversible
processes, i.e., $q_{\text{out}}=0$, and can thus only be attained
by infinitely slow driving, corresponding to vanishing power. A
more relevant measure of efficiency is therefore for example the efficiency at
maximum power (EMP), i.e.,
\begin{eqnarray}
\eta^{\text{emp}}=\frac{p_{\text{max}}}{\epsilon_{\text{in}}}.
\label{emp}
\end{eqnarray}

In a recent paper \cite{Fogedby04a} we analyzed the motor model by Mogilner et al. 
from a purely analytical point of view and derived
explicit expressions for the motion of the motor and the velocity-load relationship. 
In the present paper we return to the analytical solution of the
Mogilner model and focus on the energetics and efficiency of the model; these
aspects were not considered in \cite{Fogedby04a}. 

The paper is organized in the following manner. In Sec.
\ref{model} we review the Mogilner model and the analytical
solution. In Sec. \ref{energy} we address the energetics and
derive an expression for the efficiency. In Sec. \ref{discussion}
we discuss the efficiency of the motor in detail. Section  \ref{summary}
is devoted to a brief summary. 
\section{\label{model} Model}
We consider a minimal power stroke model of a motor molecule or
nano engine operating in an isothermal environment. We assume that
the motor is driven by the ATP hydrolysis. During a cycle an ATP
molecule is attached to the motor releasing the free energy
$\Delta\mu$. The motor is composed of two sections or heads, an
active head and a passive head. During a conformational change the
positions of the heads change. Denoting the equilibrium distance between the
heads by $L$ we assume that the motor can exist in only two
geometrical conformations, corresponding to the values $L_r$ and
$L_s$. During a power stroke induced by the attachment of an ATP
molecule to the motor and the subsequent hydrolysis of ATP  the 
motor undergoes a conformational change from a relaxed
state R with geometrical parameter $L_r$ to a strained state S
with parameter $L_s$. We, moreover, characterize the elastic
states of the motor by two spring constants $k_r$ and $k_s$,
referring to the relaxed and strained states, respectively.

The state of the motor is driven by the ATP hydrolysis. This two
step stochastic process is modeled by two coupled master
equations for the respective probabilities $P_s$ and $P_r$,
\begin{eqnarray}
&&\frac{dP_s}{dt}=g_sP_r-g_rP_s, \label{mas1}
\\
&&\frac{dP_r}{dt}=g_rP_s-g_sP_r. \label{mas2}
\end{eqnarray}
Here $g_r$ is the transition rate from the strained state S to the
relaxed state R and $g_s$ the rate from R to S. In the stationary
state we have
\begin{eqnarray}
&&P_s=\frac{g_s}{g_s+g_r}, \label{stat1}
\\
&&P_r=\frac{g_r}{g_s+g_r}. \label{stat2}
\end{eqnarray}
Simple arguments show that in order to obtain the motor property,
i.e., the motion of the motor in the absence of a load, at least two 
internal coordinates $x$ and $y$ are required, accounting for the 
conformal changes. The motor, moreover, has to interact
with a track. In the Mogilner model the interaction with the track
is modeled by a time dependent friction $\zeta(t)$ which is
synchronized with the ATP hydrolysis and the associated conformal
transitions assume two values. When the active head at coordinate
$x$ is in contact with the track during a cycle we assume that
$\zeta(t)=\zeta_p$, where $\zeta_p$ is the so-called protein
friction. During the phase where the motor is detached from the
track we assume that $\zeta(t)=\zeta_v$, where $\zeta_v$ is
the friction due to the solvent. This friction is of the order of the 
Stokes value $6\pi R\eta$; here $R$ is the size of
the motor and $\eta$ the viscosity of the medium. This scenario,
discussed in detail in \cite{Fogedby04a} is described by the
coupled equations of motion
\begin{eqnarray}
&&\zeta(t)\frac{dx}{dt}=-\frac{dU}{dx}-f, \label{eq1}
\\
&&\zeta_v\frac{dy}{dt}=-\frac{dU}{dy}, \label{eq2}
\end{eqnarray}
where $U$ is the time dependent harmonic potential
\begin{eqnarray}
U(x,y,t)=\frac{k(t)}{4}[y-x-L(t)]^2. \label{pot}
\end{eqnarray}
The time dependence of $k(t)$ and $L(t)$ is given by the stochastic switches 
of the motor between the internal states, characterized by the
master equations (\ref{mas1}) and (\ref{mas2}). 
In (\ref{eq1}) and (\ref{eq2}) we are considering the
over damped case relevant in biology and, moreover, ignore the
thermal noise which takes place on a much faster time scale than
the conformational changes in the motor. In (\ref{eq1}) we have
included a load force $f$ acting on the active head at coordinate $x$.

The procedure for solving the motor problem is straightforward.
Assuming a general time dependence of $\zeta(t)$, $k(t)$, and $L(t)$
the linear equations (\ref{eq1}) and (\ref{eq2}) with potential (\ref{pot}) are
readily solved analytically. In order to compute distributions and
averages the calculation is completed by averaging over $\zeta(t)$,
$k(t)$, and $L(t)$ according to the master equations (\ref{mas1})
and (\ref{mas2}). Regarding the stationary mean velocity $v$ both
with and without load $f$, this calculation using residence time
distributions was carried out in \cite{Fogedby04a}.

\begin{figure}
\includegraphics[width=1.0\hsize]{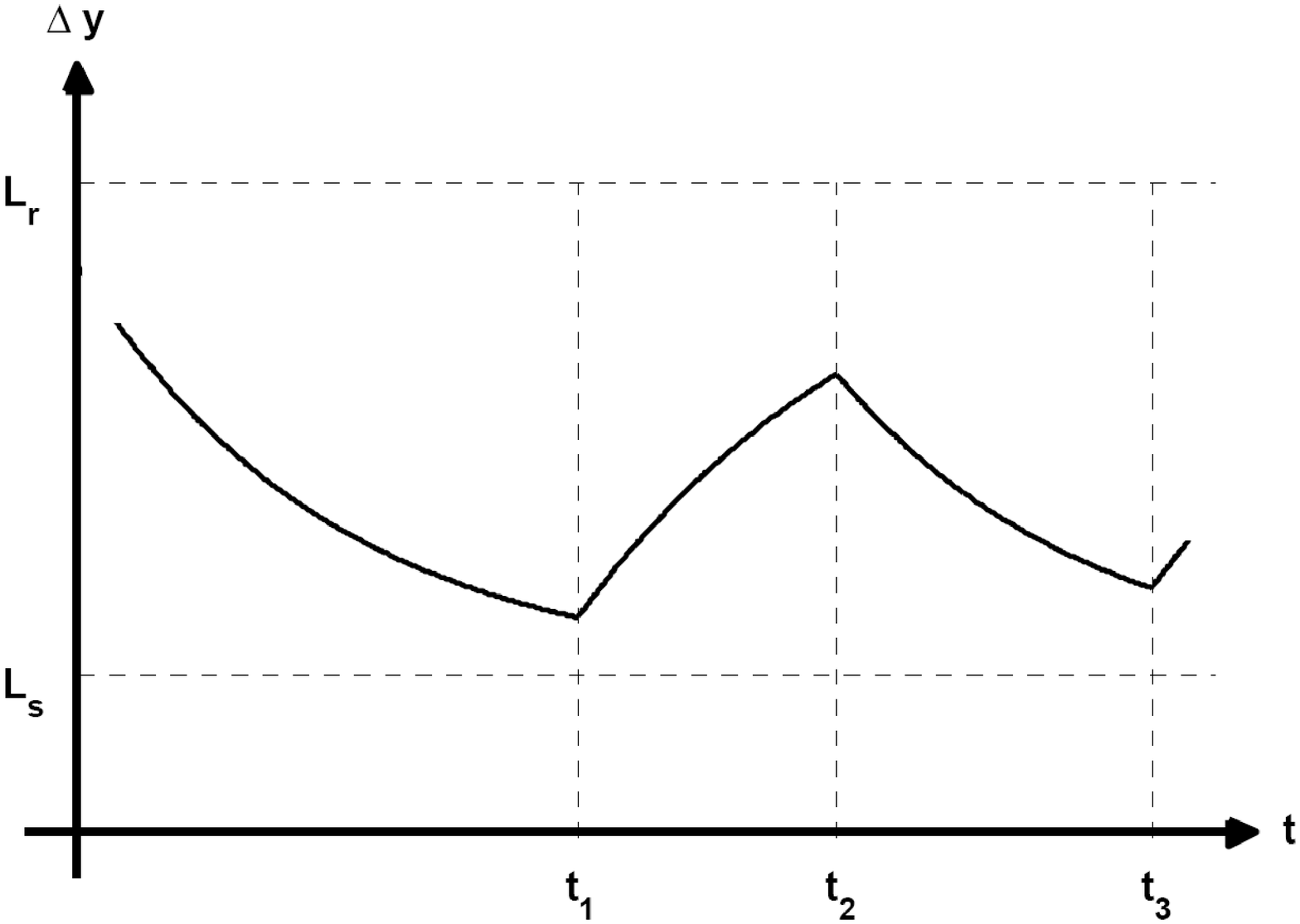}
\caption{The relative coordinate $\Delta y$ as a function of time $t$ for a concrete realization of
the time dependent parameters. At the time instants $t_n$ there are 
transitions between the relaxed and strained states driven by the master 
equations (arbitrary units).} \label{fig1}
\end{figure}

For a concrete realization of the motor parameters $\zeta(t)$,
$k(t)$, and $L(t)$, i.e., $\zeta(t)=\{\zeta_p,\zeta_v\}$,
$k(t)=\{k_r,k_s\}$, and $L(t)=\{L_r,L_s\}$, and introducing the
relative position $\Delta y=y-x$ and the damping parameter
$\Gamma=(k/2)(1/\zeta+1/\zeta_v)$ (note, in Ref. \onlinecite{Fogedby04a} 
this quantity was denoted $\dot\gamma$) we have
\begin{eqnarray}
\frac{d\Delta y}{dt}= -\Gamma\Delta y+\Gamma L+\frac{f}{\zeta},
\label{eq3}
\end{eqnarray}
with solution
\begin{eqnarray}
\Delta y= L+\frac{f}{\zeta\Gamma}+\left(\Delta
y_0-L-\frac{f}{\zeta\Gamma}\right)\exp(-\Gamma t); \label{sol}
\end{eqnarray}
here $\Delta y_0$ is the initial value at $t=0$. At long times
$t\gg 1/\Gamma$ we obtain
\begin{eqnarray}
&&(\Delta y)_r=L_r+\frac{2f}{k_r(1+\zeta_p/\zeta_v)}, \label{yr}
\\
&&(\Delta y)_s=L_s+\frac{f}{k_s}, \label{ys}
\end{eqnarray}
and for the mean separation
\begin{eqnarray}
\langle\Delta y\rangle=P_r(\Delta y)_r+P_s(\Delta y)_s,
\label{my}
\end{eqnarray}
where $P_r$ and $P_s$ are given by (\ref{stat1}) and
(\ref{stat2}). Note that since $d\Delta y/dt=0$ for $t\gg
1/\Gamma$ it follows that $\langle v_x\rangle=\langle v_y\rangle$,
i.e., the two heads move together on the average. However, as
shown in \cite{Fogedby04a} the mean velocity $v=(1/2)(\langle
v_x\rangle+\langle v_y\rangle)$ is non vanishing even for $f=0$,
establishing the motor property. As an illustration we have in Fig.~\ref{fig1}
sketched the behavior of $\Delta y$ as a function
of $t$ for $f=0$. The time instants $t_1$, $t_2$, etc. indicate when
we have transitions between the relaxed and strained states. 
\section{\label{energy} Energetics and efficiency}
Here we turn to the main issues of the present paper, namely the
energetics and efficiency of the ATP driven motor model discussed
in the previous section.
\subsection{Energetics}
\begin{figure}
\includegraphics[width=1.0\hsize]{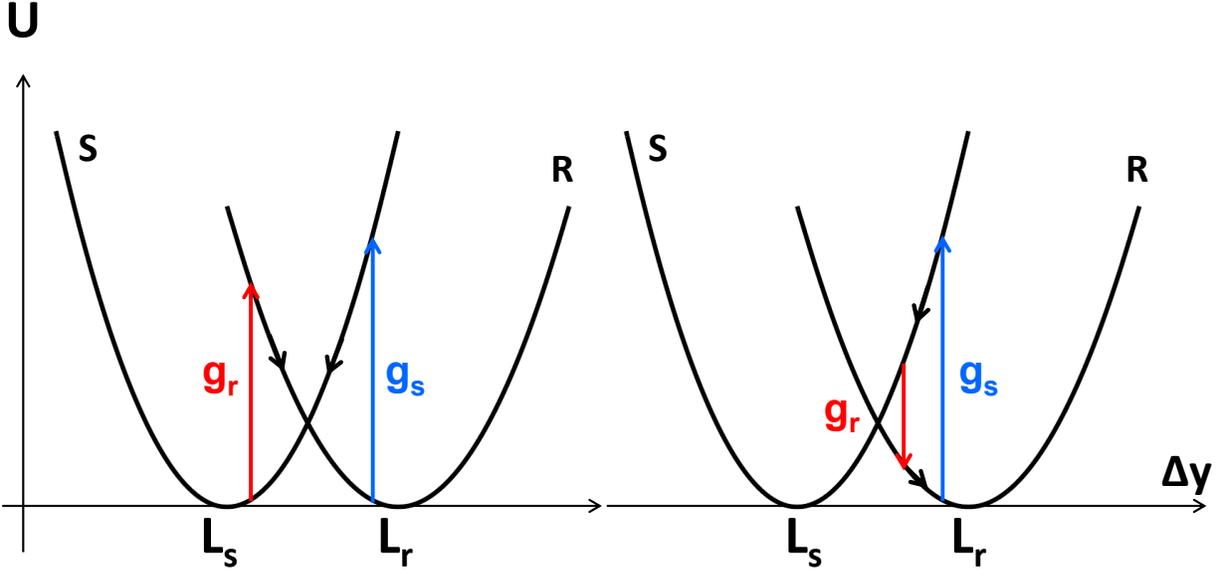}
\caption{The dissipation mechanism: In state S the motor undergoes
a transition to state R with rate $g_r$ and, subsequently, relaxes until a new transition 
with rate $g_s$ brings the motor back to state S. The momentary gain in potential 
energy is relaxed due to dissipation as indicated by the arrows. $\Delta y$ is the relative
coordinate and $U$ the potential in  (\ref{pot2})  for $f=0$ (arbitrary units). }. \label{fig2}
\end{figure}

Subject to the ATP hydrolysis energy is imparted to the motor,
temporarily stored in the spring, dissipated owing to the
friction, and performing work on the environment. Energy
conservation is expressed in (\ref{econ}),
i.e., $\epsilon_{\text{in}}=p+q_{\text{out}}$, where $\epsilon_{\text{in}}$ is
the energy input rate, $q_{\text{out}}$ the dissipation rate, and
$p$ the rate of work performed on the surroundings.
It is instructive to sketch the energy flow. Expressing the over
damped equation of motion (\ref{eq3}) in the form
\begin{eqnarray}
&&\frac{d\Delta y}{dt}=-\Gamma\frac{d\widetilde{U}}{d\Delta y},
\label{eq4}
\\
&&\widetilde{U}=\frac{1}{2}\left(\Delta
y-L-\frac{f}{\Gamma\zeta}\right)^2. \label{pot2}
\end{eqnarray}
The main effect is due to the conformational changes of
the rest length $L$. At a given time instant a transition with
rate $g_r$ excites the motor from the S state to the R state
increasing the potential energy. Subsequently, the energy is
dissipated by friction until a new transition from state R to
state S with rate $g_s$ again increases the potential energy.
In Fig.~\ref{fig2} we have for $f=0$ illustrated the decay
mechanism for  two
possible scenarios. By inspection it is clear that over a cycle the change in
potential energy $\Delta\widetilde{U}>0$, i.e., a positive
dissipation consistent with the second law of thermodynamics.

The energy conservation is also easily extracted from the equations
of motion (\ref{eq1}) and (\ref{eq2}). The dissipation force on the active
head, i.e., the head in contact with the track, is $-\zeta(t)dx/dt$,
whereas the dissipation force on the passive or free head is $-\zeta_v dy/dt$.
Consequently, the mean rate of heat dissipation is given by
\begin{eqnarray}
q_{\text{out}} = \langle\zeta(t)(dx/dt)^2\rangle
+\langle\zeta_v(dy/dt)^2\rangle. \label{dis}
\end{eqnarray}
Multiplying (\ref{eq1}) and (\ref{eq2}) by $dx/dt$ and $dy/dt$,
adding, and averaging over the power stroke cycles we obtain
$q_{\text{out}} = -p-(1/2)\langle k(\Delta y-L)d\Delta
y/dt\rangle$. Here the last term is the rate of work performed on
the head and tails of the motor. Since the motor does not
accumulate energy we can identify the input heat flux
$\epsilon_{\text{in}} =-(1/2)\langle k(\Delta y-L)d\Delta
y/dt\rangle$, in accordance with the energy balance (\ref{econ}).

The mean power given by $p=fv$, where $v$ is the mean velocity of
the  motor, is readily accessible. The mean velocity was computed
in \cite{Fogedby04a} and is given by
\begin{eqnarray}
v=v^0-\mu f, \label{vel}
\end{eqnarray}
where $v^0$ is the motor velocity in the absence of a load and
$\mu$ the mobility. The velocity $v^0$ can be written in the form
\begin{eqnarray}
&&v^0=v_h^0~\frac{\zeta_p-\zeta_v}{\zeta_p+\zeta_v}\frac{1}{1+q_r+q_s},
\label{vel1}
\\
&&v_h^0=\frac{g_sg_r}{g_r+g_s}\frac{L_r-L_s}{2}=\frac{L_r-L_s}{2(\langle t\rangle_r+\langle t\rangle_s)}. \label{vel2}
\end{eqnarray}
Here $v_h^0$ is a heuristic expression for the velocity solely
based on the conformations $L_r$ and $L_s$ and the residence
times $\langle t\rangle_s=1/g_r$ and $\langle t\rangle_r=1/g_s$,
see \cite{Fogedby04a}.
Taking $v^0=v_h^0$ neglects the internal dynamics of the motor, see
\cite{Mogilner98}. In the correction factor the dimensionless parameters
\begin{eqnarray}
&&q_s=\frac{g_r}{\Gamma_s},~~\text{where}~~\Gamma_s=\frac{k_s}{\zeta_v}, \label{qs}
\\
&&q_r=\frac{g_s}{\Gamma_r},~~\text{where}~~\Gamma_r=\frac{k_r}{2}\left[\frac{1}{\zeta_v}+
\frac{1}{\zeta_p}\right],
\label{qr}
\end{eqnarray}
express the ratios between the spring relaxation times
$\Gamma_s^{-1}$ and $\Gamma_r^{-1}$ and the residence times. 
We note that the coasting velocity $v^0$
vanishes for $L_r=L_s$, corresponding to the absence of conformational changes, and for
$\zeta_p=\zeta_v$, i.e., in the absence of a track providing protein friction; for
further discussion see \cite{Fogedby04a}.

The mobility $\mu$ determines the response of the motor to the
load force $f$. In the absence of fluctuations, i.e., the case of
constant $k$, constant $L$, and $\zeta(t)=\zeta_p$, we have from
(\ref{eq1}) and (\ref{eq2}) $\mu=1/(\zeta_p+\zeta_v)$, see
\cite{Mogilner98,Fogedby04a}. In the general case the mobility
depends on the model parameters,
\begin{equation}
\mu= \frac{[P_r\zeta_v+P_s\zeta_p]
[2q_s+(1+\zeta_v/\zeta_p)q_r]+(1+P_r)\zeta_v+ P_s\zeta_p }{
2\zeta_v(\zeta_v+\zeta_p)(1+q_s+q_r)}\label{mob}.
\end{equation}
Finally, for the mean power we have
\begin{equation}
p=(v^0-\mu f)f, \label{pow}
\end{equation}
where $v^0$ and $\mu$ are given by (\ref{vel1}), (\ref{vel2}), and
(\ref{mob}).

The remaining issue in order to establish the energetics of the
motor is the evaluation of $\epsilon_{\text{in}}$ or, alternatively,
$q_{\text{out}}$. It turns out to be most convenient to consider
the heat dissipation rate $q_{\text{out}}$ given by (\ref{dis}).
The evaluations of $q_{\text{out}}$ with or without load can be carried out using
the waiting time distribution method employed in \cite{Fogedby04a}. 
However, in the present paper we shall carry out the evaluation using transition
probabilities and a more automatic many body scheme. This still rather complex
machinery is deferred  to Appendix \ref{app}.
In the load free case, $f=0$, we
find (see Eq. (\ref{qoutf0}) in the Appendix): 
\begin{eqnarray}
q_{\text{out}}^0=\frac{(L_r-L_s)^2g_rg_s}{2(g_r+g_s)}\times
\frac{k_r(1+q_r)+k_s(1+q_s)}{(q_r+q_s+1)(q_r+q_s+2)}~~.\label{qout2}
\end{eqnarray}

In the case of a finite load the output heat rate also depends on $f$ and has a
complicated form derived in the Appendix. In terms of
the appropriate mean values of dynamic variables, i.e.,  $\langle A\rangle=A_rP_r+A_sP_s$, it reads
\begin{eqnarray}
q_{\text{out}}=&&\frac{a^2}{2}\langle bL^2\Gamma k\rangle-
\frac{fa}{\zeta_v}\langle L k\rangle +\frac{f^2}{\zeta_v}
-\frac{a^2}{c+\langle\Gamma\rangle}(c\langle bL^2\Gamma k\rangle+
\langle bL\Gamma\rangle\langle L\Gamma k\rangle) \nonumber \\
&& +\frac{fa}{\zeta_v(c+\langle\Gamma\rangle)}(c\langle bL k\rangle+
\langle bL\Gamma\rangle\langle k\rangle)  +
\frac{a^2}{2(c+\langle\Gamma\rangle)(2c+\langle\Gamma\rangle)} \times \nonumber \\
&&\left[\langle bL\Gamma\rangle^2\langle\Gamma k\rangle +
c\langle b^2L^2\Gamma\rangle\langle\Gamma k\rangle+
2c\langle bL\Gamma\rangle\langle bL\Gamma k\rangle +
 2c^2\langle b^2L^2\Gamma k\rangle \right],
\label{qoutav1}
\end{eqnarray}
where
\begin{eqnarray}
&&a=1-2f\frac{k_r-k_s}{k_rk_s(L_r-L_s)},\\
&&b=1-\frac{f}{a\zeta_v}(L\Gamma)^{-1},\\
&&c=\frac{\Gamma_r\Gamma_s}{g_r+g_s}.
\label{abc}
\end{eqnarray}
\section{\label{discussion} Discussion}
Here we proceed to discuss the energetics and efficiency of the motor. First, let us focus on the 
energy bookkeeping. The motor is driven by the attachment of ATP and the subsequent 
hydrolysis to ADP and P. 
This chemical reaction triggers the motor as expressed by the rate $g_s$. During a cycle
the system absorbs the energy $\Delta\mu$  arising from the ATP hydrolysis. Consequently,
the rate of energy input is $\Delta\mu g_s$, where $g_s$ is the transition rate from
the relaxed motor state $R$ to the strained motor state $S$. The amount of energy which the motor
can absorb per cycle depends on the actual state of the motor at the moment
the hydrolysis takes place, see Figs. \ref{fig1} and \ref{fig2}. Consequently,  we can define the intermediate
ATP related efficiency
\begin{eqnarray}
\eta^{\text{atp}}=\frac{\epsilon_{\text{in}}}{\Delta\mu g_s},
\label{eff2}
\end{eqnarray}
where $\epsilon_{\text{in}}$ is the rate of input energy recovered by the motor. The size
of $\eta^{\text{atp}}$, while not exceeding $\eta^{\text{atp}}=1$, depends on the 
frictional energy rate, $\Delta\mu g_s-\epsilon_{\text{in}}$, absorbed by  
degrees of freedom not included in the present motor model. In that sense $\eta^{\text{atp}}$
is a "fudge" parameter which can only be estimated qualitatively. The efficiency specifically 
associated with the motor model is given by 
$\eta^{\text{motor}}=p/\epsilon_{\text{in}}$, where $p$ as given by Eq. (\ref{pow}), expresses the work exerted on the environment.
This efficiency is given by 
\begin{eqnarray}
\eta^{\text{motor}}=\frac{(v^0-\mu f)f}{q_{\text{out}}+(v^0-\mu
f)f}. \label{eff1}
\end{eqnarray}
The total efficiency, $\eta_2$, relating the power $p$ to the
ATP energy input rate is thus given by the product of the two efficiencies: 
\begin{eqnarray}
\eta_2=\eta^{\text{atp}}\eta^{\text{motor}}. \label{effprod}
\end{eqnarray}
In discussing the motor efficiency $\eta^{\text{motor}}$ it is useful to first as a reference use the heuristic expressions 
by Mogilner et al.  \cite{Mogilner98}. Neglecting the internal dynamics of the motor
the velocity for $f=0$ is given by $v^0_h=(L_r-L_s)g_sg_r/(g_r+g_s)$ in (\ref{vel2}). 
In the presence of a load $f$ the velocity is to a first approximation reduced by
$f/\zeta_p$, where $\zeta_p$ is the protein friction, i.e., $v_r=v_r^0-f/\zeta_p$, with
mobility $\mu=1/\zeta_p$. A more detailed analysis based on (\ref{eq1}) and (\ref{eq2}), see also
 \cite{Fogedby04a}, yields $\mu=1/(\zeta_v+\zeta_p)$ in the relaxed state $R$ and
$\mu=1/2\zeta_v$ in the strained state $S$. Interpolating between the two states we can thus define
the effective mobility
\begin{eqnarray}
\mu_{h}=P_r\frac{1}{\zeta_p+\zeta_v}+P_s\frac{1}{2\zeta_v},
\label{mob2}
\end{eqnarray}
which agrees with (\ref{mob}) for $q_s=q_r=0$. Hence, we obtain for $v_h$
\begin{eqnarray}
v_h=v_h^0-\mu_h f.
\label{vr2}
\end{eqnarray}
For the power $p_h$ neglecting the internal dynamics we thus obtain
\begin{eqnarray}
p_h=(v_h^0-\mu_h f)f.
\label{pow2}
\end{eqnarray}
On the basis of (\ref{dis}) we can make a simple estimate for $q_{\text{out}}$. In state $R$
the friction $\zeta(t)=\zeta_p$ and we have $q_{\text{out}}\sim \zeta_p v_x^2+\zeta_v v_y^2=(\zeta_p+\zeta_v)v^2$;
in state $S$ we have $\zeta(t)=\zeta_v$ and we obtain $q_{\text{out}}\sim 2\zeta_v v^2$.
By the same interpolation as in the case of $\mu_h$ we thus obtain the estimate 
\begin{eqnarray}
q^h_{\text{out}}\sim (P_r(\zeta_p+\zeta_v)+2P_s\zeta_v)v_h^2=\zeta_h v_h^2,
\label{qest}
\end{eqnarray}
where we have introduced the effective viscosity
\begin{eqnarray}
\zeta_h=P_r(\zeta_p+\zeta_v)+2P_s\zeta_v.
\label{zeteff}
\end{eqnarray}
For the efficiency we thus obtain
\begin{eqnarray}
\eta_h=\frac{v_h f}{\zeta_h v_h^2+v_h f}.
\label{eff3}
\end{eqnarray}
In this simple approximation 
\begin{eqnarray}
\eta_h=\frac{f}{\zeta_h v_h+f}.
\label{eff4}
\end{eqnarray}
For $f=0$ we have $p_h=0$ and thus $\eta_h=0$. At the stall force $f_{\text{stall}}=v_h^0/\mu_h$ we
have $v_h=0$ and the efficiency $\eta_h=1$.
The power $p_h$ has a maximum for $f=v_h^0/2\mu_h$ and we obtain the efficiency at maximum
power (emp)
\begin{eqnarray}
\eta^{\text{emp}}_h=\frac{1}{\mu_h\zeta_h+1}.
\label{empp}
\end{eqnarray}
Inserting biological parameters from Mogilner et al.  \cite{Mogilner98}, i.e., $\zeta_p=50\times 10^{-6}~\text{pNs/nm}$,
$\zeta_v= 10^{-6}~\text{pNs/nm}$, $g_r=10^3~\text{s}^{-1}$,  $g_s=10^3~\text{s}^{-1}$, $P_r=0.5$, and $P_s=0.5$, we obtain
$\mu_h=2.6\times 10^5~\text{nm/pNs}$ and $\zeta_h=2.6\times 10^{-5}~\text{pNs/nm}$, yielding the efficiency at maximum
power $\eta^{\text{emp}}_h=0.13$.

The next issue is to examine how the above heuristic expression for the motor efficiency depends on the intrinsic motor parameters.
For the mobility $\mu$, entering in the expression for the velocity and the power, we have from (\ref{mob}), see also
\cite{Fogedby04a}, 
\begin{eqnarray}
&&\mu=\mu_hC_\mu,
\\
&&C_\mu= \frac{1+
[2q_s+(1+\zeta_v/\zeta_p)q_r][P_r\zeta_v+P_s\zeta_p]/[P_r\zeta_v+P_s\zeta_p+\zeta_v]} 
{1+q_s+q_r},
\label{mob3}
\end{eqnarray}
where the correction factor depends on the motor parameters. Inserting 
$k_r=0.01~\text{pN/nm}$, $k_s=0.5~\text{pN/nm}$, yielding $q_r=0.2$
and $q_s=2\times 10^{-3}$, we obtain $C_\mu=0.998$, i.e., close to 1.
Likewise, for the coasting velocity we have from (\ref{vel1}) and (\ref{vel2})
\begin{eqnarray}
&&v^0=v^0_hC_v,
\\
&&C_v=\frac{\zeta_p-\zeta_v}{\zeta_p+\zeta_v}\frac{1}{1+q_r+q_s}.
\end{eqnarray}
Inserting biological parameters we find $v_h^0=0.5\times 10^4~\text{nm/s}$ and the correction
factor $C_v=0.8$, i.e. close to 1.
Since the power $p=(v^0-\mu f)f$ attains the maximum value $p_{\text{max}}=(v^0)^2/4\mu$ for 
$f_{\text{max}}=v^0/2\mu$ the central quantity determining $\eta^{\text{emp}}$ is the dissipation rate 
$q_{\text{out}}$ which has a dependence on the load force $f$. Referring to the heuristic ansatz
we set
\begin{eqnarray}
q_{\text{out}}=\zeta_h v_h^2 C_q(f).
\end{eqnarray}
The correction factor $C_q(f)$ is then determined from $q_{\text{out}}$ evaluated in Appendix \ref{app}.
With these definitions we have
\begin{eqnarray}
&&p=(v_h^0C_v-\mu_hC_\mu f)f,
\\
&&q_{\text{out}}=\zeta_h v_h^2 C_q,
\\
&&\eta=\frac{(v_h^0C_v-\mu_hC_\mu f)f}{\zeta_h v_h^2 C_q+(v_h^0C_v-\mu_hC_\mu f)f}.
\end{eqnarray}
Maximum power $p_{\text{max}}=(v_h^0 C_v)^2/4\mu_hC_\mu$ is attained for 
$ f_{\text{max}}=v_h^0C_v/2\mu_hC_\mu$, and we obtain the efficiency at
maximum power
\begin{eqnarray}
\eta^{\text{emp}}=\frac{(v_h^0 C_v)^2/4\mu_hC_\mu}{\zeta_h v_h^2 C_q+(v_h^0 C_v)^2/4\mu_hC_\mu}.
\end{eqnarray}
Choosing the biological parameters $L_r=40~\text{nm}$, $L_s=20~\text{nm}$, we obtain
$p_{\text{max}}=24~\text{pN}\cdot\text{nm/s}$ and $f_{\text{max}}=9.6\times 10^{-3}~\text{pN}$.
Moreover, from (\ref{qest}) $q_{\text{out}}^h=\zeta_h v_h^2= 1.6\times 10^2~\text{pN}\cdot\text{nm/s}$.
From   Eq. (\ref{qoutav1})   we obtain for $f=f_{\text{max}}$ the dissipation rate 
$q_{\text{out}}= 1.94\times 10^4~\text{pN}\cdot\text{nm/s}$, i.e., $C_q=1.2\times 10^2$, and the
maximum efficiency drops to $\eta^{\text{emp}} =1.2\times 10^{-3}$, i.e about a per  mille.
Thus, we find that including the internal dynamics of the motor $\eta^{\text{emp}}$ is reduced from about
10\% in the heuristic Mogilner case to about 1 per mille in the full motor case. The dissipation associated 
with the motor activity acts as a bottle neck reducing $\eta^{\text{emp}}$. Based on the expressions in 
Appendix \ref{app} we have in Fig.~\ref{fig3} presented a 3D plot of $\eta^{\text{emp}}$ as a function of the dimensionless
parameters $q_s$ and $q_r$ characterizing the motor dynamics. We note that for
$q_s$ of order 100 and $q_r$ of order 1000 $\eta^{\text{emp}}$ reaches a  plateau at around
an $\eta^{\text{emp}}$ of about 10\%. In Fig.~\ref{fig4} we have depicted the corresponding contour plot of
$\eta^{\text{emp}}$  as function of $q_r$ and $q_s$. Here the plateau in $\eta^{\text{emp}}$  attained for large $q_r$ and $q_s$
is clearly discernible. 

In the limit of large $q_{r,s}=(\Gamma_{r,s}\langle t\rangle_{r,s})^{-1}$ 
we have $\langle t\rangle_{r,s}\ll\Gamma_{r,s}^{-1}$, where $\langle t\rangle_{r,s}$ are the residence
times and $\Gamma_{r,s}^{-1}$ the spring relaxation times. Consequently,
the motor switches rapidly between its internal states and the fluctuations about the
internal length $L(t)$ are small. Based on this approximation one can derive approximate
expressions for $q_{\text{out}}$ and $\eta^{\text{emp}}$. The expressions are lengthy and the analysis
is therefore deferred to the Appendix. Among other results we find that scaling the
residence times by a common factor yields invariant expression for $p_{\text{max}}$ and 
$q_{\text{out}}$. This scaling behavior accounts for the approximately linear contours of constant $\eta^{\text{emp}}$
at large $q_r$ and $q_s$, depicted 
in Fig.~\ref{fig4}. In Fig.~\ref{fig5} we depict in a 3D plot the EMP, denoted $\eta^{\text{app}}$,
showing a strong agreement  with the exact result in Fig.~\ref{fig3}. 
The fact that the EMP is as low as a few per mille at physical motor parameters reflects the opposite 
limit where the residence times are larger than the spring relaxation times.
In this case the motor spends a long time in the passive mode, where the internal spring is fully relaxed, 
waiting for the next conformational change to occur. In this situation all that happens is that the load 
force pulls the motor backwards contributing negatively to $\eta^{\text{emp}}$. 

Since the efficiency
evaluated here applies to the motor {\it per se}  it is instructive to consider the efficiency relative to the 
burning of ATP, $\eta_2$, defined by (\ref{effprod}). 
In  Fig.~\ref{fig6} we have depicted $\eta_2$ at maximum power.

\section{\label{summary} Summary}
In this paper we have in some detail discussed the efficiency of the Mogilner motor
model based on protein friction \cite{Mogilner98}, extending the previous analytical 
findings in \cite{Fogedby04a} to include the energetics and efficiency. In the process
we have developed a many body scheme for the evaluation of dissipation rates
applied to the interplay between two coupled over damped equations of motion
and a two step Markov process described by coupled master equations.
Due to the strong dissipation concurrent with the motor operation we find in the
biological regime an $\eta^{\text{emp}}$ of the order of 1 per mille. The  dissipation associated
with the medium and the protein friction acts as an effective bottle neck in the transfer
of energy from the ATP hydrolysis to the power exerted on the environment.
We stress that our analysis applies to the efficiency of the motor model {\it per se}
and not the assembly including the attachment and hydrolysis of ATP. Including
this feature would require an extension of the model; this has not been attempted
in the present context.

In recent works an efficiency at maximum power, $\eta^{\text{emp}}$, approaching 100\%
has been reported, in marked 
contrast to the present results where  $\eta^{\text{emp}}$ is of order 10\% by
a heuristic estimate and furthermore reduced to about 1 per mille when
the motor mechanics is included properly. Refraining from a detailed discussion, 
see incidentally \cite{Fogedby04a},
it is common to all the studies cited above that the motor dynamics is not treated explicitly.
The motor molecule driven by ATP hydrolysis and interacting with a track
is described schematically either by a power stroke model or a
Brownian racket model. These studies throw light on the fundamental energy
transfer in a molecular motor and can due to their schematic nature yield large $\eta^{\text{emp}}$.
In the present study the motor is modeled explicitly, the price paid, however, is
a low $\eta^{\text{emp}}$ due to dissipation and an energy "bottle neck" effect.

\appendix
\section{\label{app} Evaluation of $q_{\text{out}}$}
Unlike the method used in \cite{Fogedby04a} where the 
evaluations of the mean velocity $v^0$ and the mobility $\mu$ are based on waiting time 
distributions, we here develop a technique directly based on transition probabilities 
given by the master equations (\ref{mas1}) and (\ref{mas2}). To ease notation we set 
$s=1$ and $r=2$. Denoting in the following a time derivative
by a dot, the master equations (\ref{mas1}) and (\ref{mas2})
then read $\dot P_1=g_1P_2-g_2P_1$ and $\dot P_2=g_2P_1-g_1P_2$ with stationary solutions
\begin{eqnarray}
&&P_1=\frac{g_1}{g},
\label{p1}
\\
&&P_2=\frac{g_2}{g},
\label{p2}
\\
&&g=g_1+g_2.
\label{g}
\end{eqnarray}
The conditional transition probabilities $P_{nm}(t)$ from state $n=1,2$ to state $m=1,2$
satisfy the master equations \cite{Reichl98} $\dot P_{11}=g_1P_{12}-g_2P_{11}$,
$\dot P_{22}=g_2P_{21}-g_1P_{22}$, $\dot P_{12}=g_2P_{11}-g_1P_{12}$ and 
$\dot P_{21}=g_1P_{22}-g_2P_{21}$. Imposing the boundary conditions $P_{nm}(0)=\delta_{nm}$ 
and normalization condition $\sum_mP_{nm}(t)=1$ we infer the solution 
\begin{eqnarray}
P_{nm}(t)=A_{nm}+B_{nm}\exp(-gt),
\label{sol1}
\end{eqnarray}
where we have introduced the matrices
\begin{eqnarray}
A=
\left(
\begin{array}{cc}
P_1 & P_2  \\
P_1 & P_2  \\
\end{array} \right),~~
B=
\left(
 \begin{array}{cc}
P_2&-P_2  \\
-P_1 & P_1 \\
\end{array} \right);
\label{ab}
\end{eqnarray}
note that (\ref{sol1}) satisfies the Chapman-Enskog equation \cite{Reichl98}
\begin{eqnarray}
P_{nm}(t_3-t_1)=\sum_kP_{nk}(t_2-t_1)P_{km}(t_3-t_2), ~~\text{for}~~~ t_1<t_2<t_3.
\label{ce}
\end{eqnarray}
We also require the diagonal matrix characterizing the initial stationary distribution
\begin{eqnarray}
P^0=
\left(
\begin{array}{cc}
P_1 & 0  \\
0 & P_2 \\
\end{array} \right).
\label{p0}
\end{eqnarray}
Introducing the Fourier transform
\begin{eqnarray}
P_\omega=\int_0^\infty dt\exp(i\omega t)(A+B\exp(-gt)),
\label{ft}
\end{eqnarray}
we have
\begin{eqnarray}
P_\omega=-\left(\frac{A}{i\omega}+\frac{B}{i\omega-g}\right);
\label{pom}
\end{eqnarray}
note that with this definition $\omega$ lies in the upper half plane above the real axis and  a contour
integration is performed by closing the contour in the lower half plane. Also $P_\omega$ has
poles at $\omega=0$ and $\omega=-ig$.

By quadrature the equations of motion (\ref{eq1}) and (\ref{eq2}) with potential (\ref{pot}) 
yield for $f=0$, for details see  \cite{Fogedby04a},
\begin{eqnarray}
&&v_x^0(t)=-\frac{k(t)}{2\zeta(t)}\int_0^t dt' \dot L(t')A(t,t'),
\label{vx0}
\\
&&v_y^0(t)=\frac{k(t)}{2\zeta_v}\int_0^t dt' \dot L(t')A(t,t'),
\label{vy0}
\\
&&A(t,t')=\exp\left(-\int_{t'}^td\tau\Gamma(\tau)\right),
\label{att}
\\
&&\Gamma(t)=\frac{k(t)}{2}\left(\frac{1}{\zeta(t)}+\frac{1}{\zeta_v}\right).
\label{gam}
\end{eqnarray}
Correspondingly, in the presence of a load $f$ we have
\begin{eqnarray}
&&v_x(t)=-\frac{k(t)}{2\zeta(t)\zeta_v}\int_0^t dt'\left(\tilde f(t')+\zeta_v\dot L(t')\right)A(t,t'),
\label{vx}
\\
&&v_y(t)=\frac{k(t)}{2\zeta_v^2}\int_0^t dt'\left(\tilde f(t')+\zeta_v\dot L(t')\right)A(t,t')-\frac{f}{\zeta_v},
\label{vy}
\\
&&\tilde f(t)=f\left(1-2\zeta_v\frac{\dot k(t)}{k(t)^2}\right).
\label{ftil}
\end{eqnarray}
%
\subsection{\label{app1} The mean heat for $f=0$}
We first turn to the evaluation of the average heat  $q_{\text{out}}^0$ for $f=0$ using the above scheme. 
Inserting (\ref{vx0}) and  (\ref{vy0}) in (\ref{dis}) we obtain using (\ref{gam}) 
\begin{eqnarray}
q_{\text{out}}^0(t)=\frac{1}{2}\int_0^t dt'\int_0^t dt''\langle\dot L(t')A(t,t')\dot L(t'')A(t,t'')\Gamma(t)k(t)\rangle.
\label{q0}
\end{eqnarray}
First, carrying out a partial integration, noting that the initial value $L(0)$ will not contribute in the long time
limit, we have, using $dA(t,t')/dt'=\Gamma(t')A(t,t')$,
\begin{eqnarray}
q_{\text{out}}^0(t)=\frac{1}{2}\langle\left(L(t)-\int_0^t dt'L(t')\Gamma(t')A(t,t')\right)^2\Gamma(t)k(t)\rangle.
\label{q1}
\end{eqnarray}
Expanding we have
\begin{eqnarray}
q_{\text{out}}^0(t)=A(t)+B(t)+C(t),
\label{q2}
\end{eqnarray}
where
\begin{eqnarray}
&&A(t)=\frac{1}{2}\langle L(t)^2\Gamma(t)k(t)\rangle,
\label{a}
\\
&&B(t)=-\int_0^t dt' \langle L(t')\Gamma(t')A(t,t')L(t)\Gamma(t)k(t)\rangle,
\label{bb}
\\
&&C(t)=\frac{1}{2}\int_0^t dt' \int_0^t dt'' \langle L(t')\Gamma(t')A(t,t')
L(t'')\Gamma(t'')A(t,t''))\Gamma(t)k(t)\rangle.
\label{c1}
\end{eqnarray}
Inserting (\ref{p0}) and (\ref{pom}) we obtain in Fourier space
\begin{eqnarray}
A_\omega = \frac{1}{2}\sum_{nm}[P^0P_\omega L^2\Gamma k]_{nm};
\label{a1}
\end{eqnarray}
we note that $L$, $ \Gamma$, and $k$ are diagonal matrices and that
the sum is performed over all matrix elements.

The evaluation of $B$ and $C$ requires more analysis. In order to invoke
the causal structure of the calculation we time order the expansion of $A(t,t')$,
i.e.,
\begin{eqnarray}
A(t,t')=\sum_{p=0}\frac{(-1)^p}{p!}\left(\int_{t'}^t d\tau\Gamma(\tau)\right)^p=
\sum_{p=0}(-1)^pT\left(\int_{t'}^t d\tau\Gamma(\tau)\right)^p,
\label{att1}
\end{eqnarray}
where the time ordering $T$ is defined according to, see ref. \cite{Das93},
\begin{eqnarray}
T\left(\int_{t'}^t d\tau\Gamma(\tau)\right)^p=
\int_{t'}^t dt_1\int_{t_1}^t dt_2\cdots\int_{t_{p-1}}^t dt_p\Gamma(t_1)\Gamma(t_2)\cdots\Gamma(t_p).
\label{to}
\end{eqnarray}
Considering first $B(t)$ we have
\begin{eqnarray}
B(t)=-\int_0^t dt'\sum_{p=0}(-1)^p\langle L(t')\Gamma(t')T\left(\int_{t'}^t d\tau\Gamma(\tau)\right)^pL(t)\Gamma(t)k(t)\rangle,
\label{b1}
\end{eqnarray}
and in Fourier space, noting that the time ordering corresponds to a convolution,
\begin{eqnarray}
B_\omega=-\sum_{p=0}(-1)^p\sum_{nm}[P^0P_\omega L\Gamma(P_\omega\Gamma)^pP_\omega\Gamma L k]_{nm}.
\label{b2}
\end{eqnarray}

In the case of $C(t)$ we must first time order the $t'$ and $t''$ integrations, noting that the integrand is
symmetric in $t'$ and $t''$, and subsequently break up $A(t,t')$ according to $A(t,t')=A(t,t'')A(t'',t')$ in order to
achieve the complete causal time ordering of the expression. We obtain in Fourier space
\begin{eqnarray}
C_\omega=\sum_{p=0}(-1)^p\sum_{q=0}(-2)^q\sum_{nm}[P^0P_\omega L\Gamma(P_\omega\Gamma)^pP_\omega\Gamma L
(P_\omega\Gamma)^qP_\omega \Gamma k]_{nm}.
\label{c2}
\end{eqnarray}
Rearranging (\ref{b2}) and (\ref{c2}) and carrying out the binomial sums we have
\begin{eqnarray}
&&B_\omega=-\sum_{nm}[P^0P_\omega L\Gamma P_\omega(1+\Gamma P_\omega)^{-1}
L\Gamma k]_{nm},
\label{b3}
\\
&&C_\omega =\sum_{nm}[P^0P_\omega L\Gamma P_\omega(1+\Gamma P_\omega)^{-1}
L\Gamma P_\omega(1+2\Gamma P_\omega)^{-1}\Gamma k]_{nm}.
\label{c3}
\end{eqnarray}
The long time behavior of $q_{\text{out}}^0$ is determined by the pole at $\omega=0$. 
Evaluating $(1+\Gamma P_\omega)^{-1}$ and $(1+2\Gamma P_\omega)^{-1}$ using
the diagonal matrix $\Gamma$ with matrix elements $\Gamma_1$ and $\Gamma_2$
and $P_\omega$ given by (\ref{pom}) it is easily seen that these factors are constant for $\omega=0$.
Consequently, setting $\Gamma P_\omega(1+\Gamma P_\omega)^{-1}=
 1-(1+\Gamma P_\omega)^{-1}$ and  $\Gamma P_\omega(1+2\Gamma P_\omega)^{-1}=
(1/2)( 1-(1+2\Gamma P_\omega)^{-1})$ the factor $P_\omega$ possessing a pole at $\omega=0$, $P_\omega\sim
-A/i\omega$, governs the long time behavior. At $\omega=0$ we obtain
\begin{eqnarray}
&&(1+\Gamma P_\omega)^{-1}|_{\omega=0}=F_1C,
\label{re1}
\\
&&(1+2\Gamma P_\omega)^{-1}|_{\omega=0}=F_2C,
\label{re2}
\\
&&F_1=\frac{1}{(P_1\Gamma_1+P_2\Gamma_2)+\Gamma_1\Gamma_2/g},
\label{f1}
\\
&&F_2=\frac{1}{(P_1\Gamma_1+P_2\Gamma_2)+2\Gamma_1\Gamma_2/g},
\label{f2}
\\
&&C=
\left(\begin{array}{cc}\Gamma_2P_2 & -\Gamma_1P_2 \\ -\Gamma_2P_1 & \Gamma_1P_1\end{array}\right),
\label{c45}
\end{eqnarray}
For $A_\omega$,  $B_\omega$, and $C_\omega$ we then obtain the pole contributions 
\begin{eqnarray}
&&A_\omega =-\frac{1}{2i\omega}\sum_{nm}[P^0AL^2\Gamma k]_{nm},
\label{a4}
\\
&&B_\omega =+\frac{1}{i\omega}\sum_{nm}[P^0AL(1-F_1C)L\Gamma k]_{nm},
\label{b4}
\\
&&C_\omega =-\frac{1}{2i\omega}\sum_{nm}[P^0AL(1-F_1C)L(1-F_2C)\Gamma k]_{nm}.
\label{c4}
\end{eqnarray}
Rewriting $1-F_1C$ and $1-F_2C$ in the form
\begin{eqnarray}
&&1-F_1C=\frac{cI+\Gamma A}{c+\langle\Gamma\rangle},
\\
&&1-F_2C=\frac{2cI+\Gamma A}{2c+\langle\Gamma\rangle},
\\
&&c=\frac{\Gamma_1\Gamma_2}{g},
\end{eqnarray}
using the identities
\begin{eqnarray}
&&AdA=A\langle d\rangle,
\\
&&\sum_{nm}[P^0Ad]_{nm}=\langle d\rangle,
\end{eqnarray}
where $d$ is a diagonal matrix, and extracting  $q_{\text{out}}^0$ from 
$q_{\text{out}}^0=\int(d\omega/2\pi)\exp(-i\omega t)(A_\omega+B_\omega+C_\omega)$
by closing the contour in the lower half plane, picking up the residue from the pole
at $\omega=0$, we obtain after further reduction
\begin{eqnarray}
q_{\text{out}}^0=&&\frac{1}{2(2c+\langle\Gamma\rangle)}
[\langle L^2\Gamma k\rangle-2\langle L\Gamma\rangle\langle L\Gamma k\rangle]+
\nonumber
\\
&&\frac{1}{2(c+\langle\Gamma\rangle)(2c+\langle\Gamma\rangle)}
[c\langle L^2\Gamma\rangle\langle\Gamma k\rangle+\langle L\Gamma\rangle^2\langle\Gamma k\rangle],
\label{qoutav}
\end{eqnarray}
or inserting
\begin{eqnarray}
q_{\text{out}}^0=(L_1-L_2)^2
\frac{g_1g_2\Gamma_1\Gamma_2(k_1\Gamma_2(g_2+\Gamma_1)+k_2\Gamma_1(g_1+\Gamma_2))}
{2(g_1+g_2)(g(P_1\Gamma_1+P_2\Gamma_2)+\Gamma_1\Gamma_2)(g(P_1\Gamma_1+P_2\Gamma_2)+2\Gamma_1\Gamma_2)}.
\label{qoutf0}
\end{eqnarray}
%
\subsection{\label{app2} The mean heat for $f\neq 0$}

We proceed to evaluate $q_{\text{out}}$ in the presence of a load following the method above.
First noting that $\dot\Gamma$ is synchronized with $\dot L$ we can express (\ref{ftil}) in the form
\begin{eqnarray}
&&\frac{\tilde f}{\zeta_v}+\dot L=\frac{f}{\zeta_v}+a\dot L,
\label{ftil2}
\\
&&a=1-2f\frac{k_1-k_2}{k_1k_2(L_1-L_2)},
\label{as}
\end{eqnarray}
and we have performing a partial integration                                                      
\begin{eqnarray}
&&v_x(t)=-a\left[L(t)+\int_0^tdt'\left(\frac{f}{a\zeta_v}-L(t)\Gamma(t)\right)A(t,t')\right]\frac{k(t)}{2\zeta(t)},
\label{vx1}
\\
&&v_y(t)=+a\left[L(t)+\int_0^tdt'\left(\frac{f}{a\zeta_v}-L(t)\Gamma(t)\right)A(t,t')\right]\frac{k(t)}{2\zeta_v}-\frac{f}{\zeta_v}.
\label{vx2}
\end{eqnarray}
Inserting in  (\ref{dis}) and setting
\begin{eqnarray}
b=1-\frac{f}{a\zeta_v}(L\Gamma)^{-1},
\label{c}
\end{eqnarray}
we obtain
\begin{eqnarray}
q_{\text{out}}(t)=\tilde A(t)+\tilde B_1(t)+\tilde B_2(t)+\tilde C(t),
\end{eqnarray}
where 
\begin{eqnarray}
&&\tilde A(t)=(a^2/2)\langle L(t)^2\Gamma(t)k(t)\rangle-(fa/\zeta_v)\langle L(t)k(t)\rangle+(f^2/\zeta_v),
\\
&&\tilde B_1(t)=-a^2\int_0^t dt'\langle b(t')L(t')\Gamma(t')A(t,t')L(t)\Gamma(t)k(t)\rangle,
\\
&&\tilde B_2(t)=(fa/\zeta_v)\int_0^t\langle b(t')L(t')\Gamma(t')A(t,t')k(t)\rangle,
\\
&&\tilde C(t)=(a^2/2)\int_0^t dt'\int_0^t dt''\langle b(t')L(t')\Gamma(t')A(t,t')b(t'')L(t'')\Gamma(t'')A(t,t'')\Gamma(t)k(t)\rangle.
\end{eqnarray}
We note that for $f=0$ we have $a=1$ and $b=1$ and we recover the case for $q_{\text{out}}^0$.
In Fourier space we obtain as above
\begin{eqnarray}
&&\tilde A_\omega=(-1/i\omega)\left((a^2/2)\sum_{nm}[P^0AL^2\Gamma k]_{nm}-(fa/\zeta_v)\sum_{nm}[P^0ALk]_{nm}+f^2/\zeta_v\right),
\\
&&\tilde B_{1\omega}=(-1/i\omega)\left(-a^2\sum_{nm}[P^0AbL(1-F_1C)L\Gamma k]_{nm}\right),
\\
&&\tilde B_{2\omega}=(-1/i\omega)\left((fa/\zeta_v)\sum_{nm}[P^0AbL(1-F_1C)k]_{nm}\right),
\\
&&\tilde C_\omega=(-1/i\omega)\left((a^2/2)\sum_{nm}[P^0AbL(1-F_1C)bL(1-F_2C)\Gamma k]_{nm}\right),
\end{eqnarray}
and for $q_{\text{out}}$
\begin{eqnarray}
q_{\text{out}}=&&(a^2/2)\sum_{nm}[P^0AL^2\Gamma k]_{nm}-(fa/\zeta_v)\sum_{nm}[P^0ALk]_{nm}+f^2/\zeta_v
\nonumber
             \\
             &&-a^2\sum_{nm}[P^0AbL(1-F_1C)L\Gamma k]_{nm}+(fa/\zeta_v)\sum_{nm}[P^0AbL(1-F_1C)k]_{nm}
             \nonumber
             \\
             &&+(a^2/2)\sum_{nm}[P^0AbL(1-F_1C)bL(1-F_2C)\Gamma k]_{nm}
\end{eqnarray}
The final reduction proceeds as in the case above for $q_{\text{out}}^0$. We obtain
\begin{eqnarray}
q_{\text{out}}=&&\frac{a^2}{2}\langle bL^2\Gamma k\rangle-\frac{fa}{\zeta_v}\langle L k\rangle +\frac{f^2}{\zeta_v}
-\frac{a^2}{c+\langle\Gamma\rangle}(c\langle bL^2\Gamma k\rangle+\langle bL\Gamma\rangle\langle L\Gamma k\rangle) \nonumber \\
&& +\frac{fa}{\zeta_v(c+\langle\Gamma\rangle)}(c\langle bL k\rangle+\langle bL\Gamma\rangle\langle k\rangle)  +
\frac{a^2}{2(c+\langle\Gamma\rangle)(2c+\langle\Gamma\rangle)} \times \nonumber \\
&&\left[\langle bL\Gamma\rangle^2\langle\Gamma k\rangle +
c\langle b^2L^2\Gamma\rangle\langle\Gamma k\rangle+2c\langle bL\Gamma\rangle\langle bL\Gamma k\rangle + 
2c^2\langle b^2L^2\Gamma k\rangle \right].
\label{qoutav21}
\end{eqnarray}
%
\subsection{\label{largeq} The large q limit, $q_r>>1$, $q_s>>1$.}
\begin{figure}
\includegraphics[width=1.0\hsize]{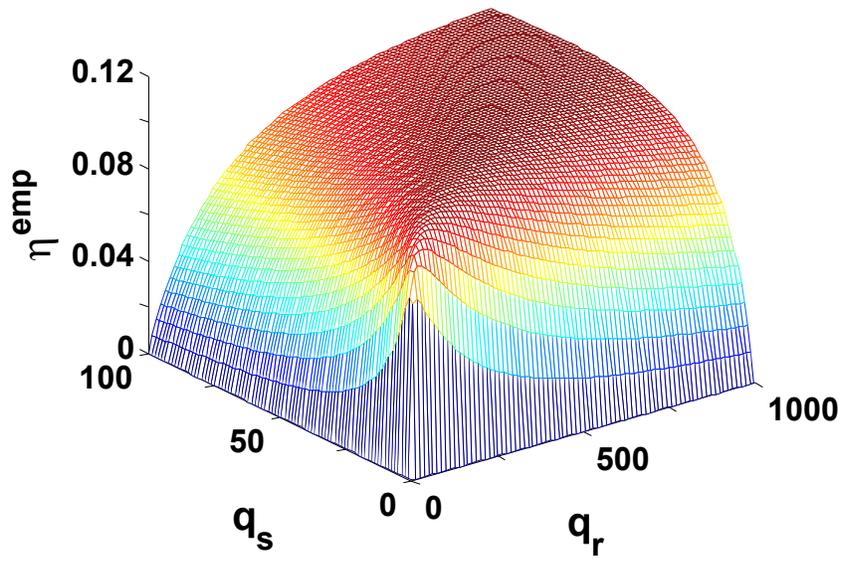}
\caption{3D plot of the exact expression for the EMP, $\eta^{\text{emp}}$, given by (\ref{qoutav1}) versus $q_r$ and $q_s$.
The range of $q_s$ is 0 to 100; the range of $q_r$ is 0 to 1000.  $\eta^{\text{emp}}$ ranges from 0 to 0.12.} \label{fig3}
\end{figure}

\begin{figure}
\includegraphics[width=1.0\hsize]{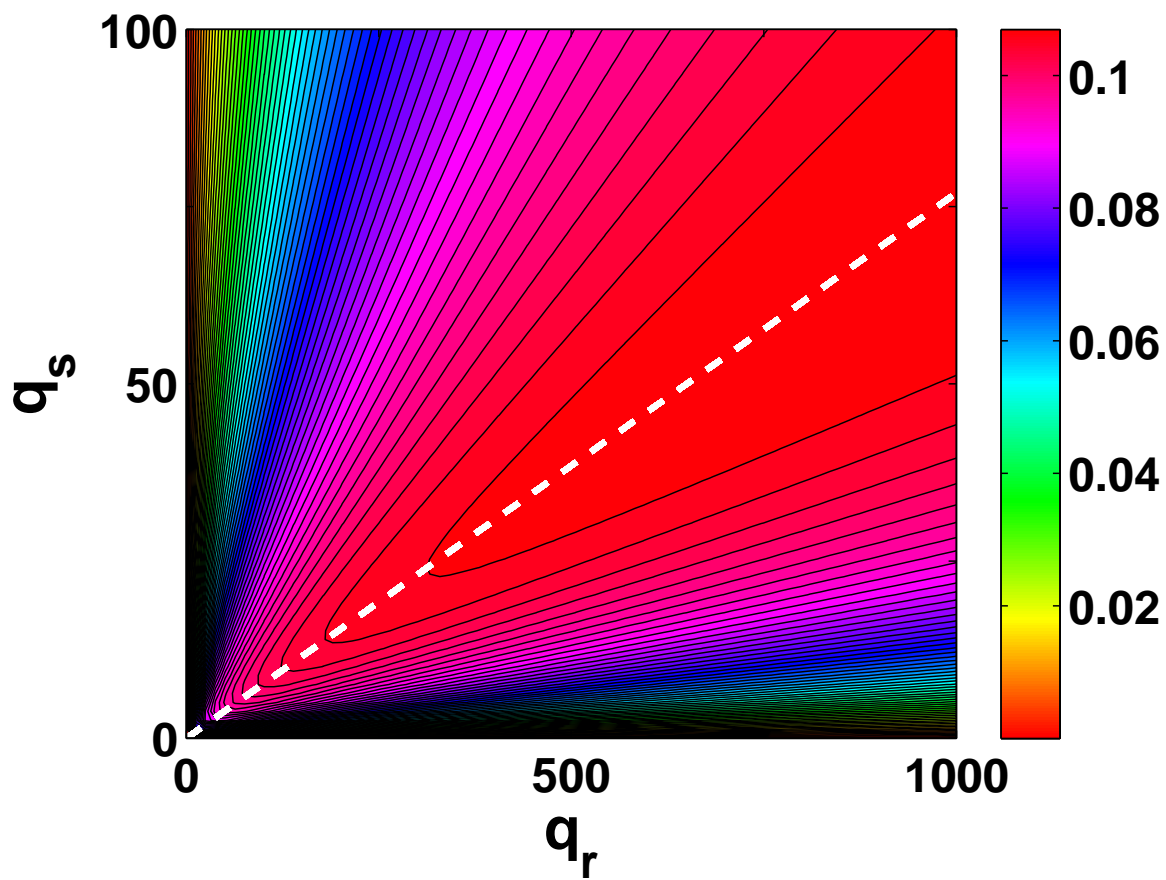}
\caption{Contour plot of the exact expression for the EMP, $\eta^{\text{emp}}$  given by  (\ref{qoutav1}), versus $q_r$ and $q_s$. 
The white dashed line corresponds to the conditions $q_r=13q_s$. The maximum efficiency $\eta^{\text{emp}}=0.12$ is attained along 
this line for large values of $q$.} \label{fig4}
\end{figure}

\begin{figure}
\includegraphics[width=1.0\hsize]{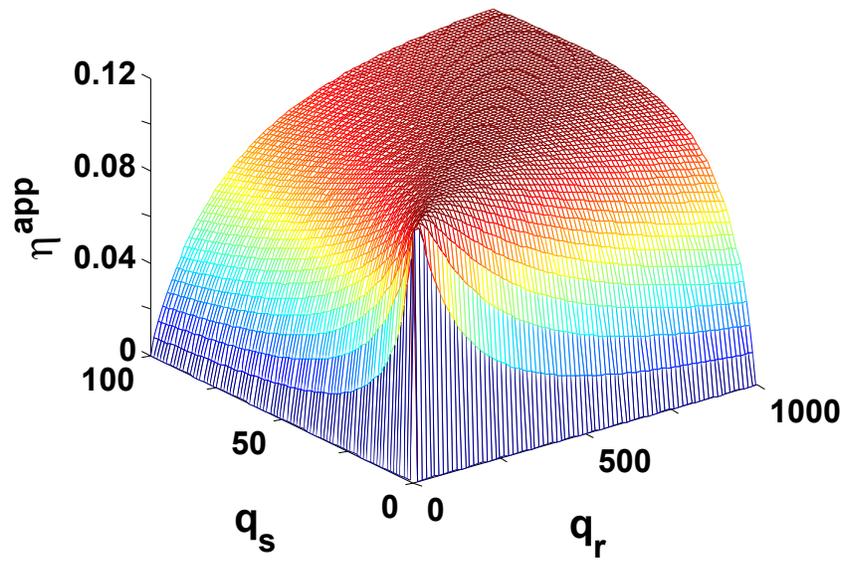}
\caption{3D plot of the approximate (large-$q$) expression for the EMP,
$\eta^{\text{app}}$,  versus $q_r$ and $q_s$, derived in Appendix A.3.} \label{fig5}
\end{figure}

\begin{figure}
\includegraphics[width=1.0\hsize]{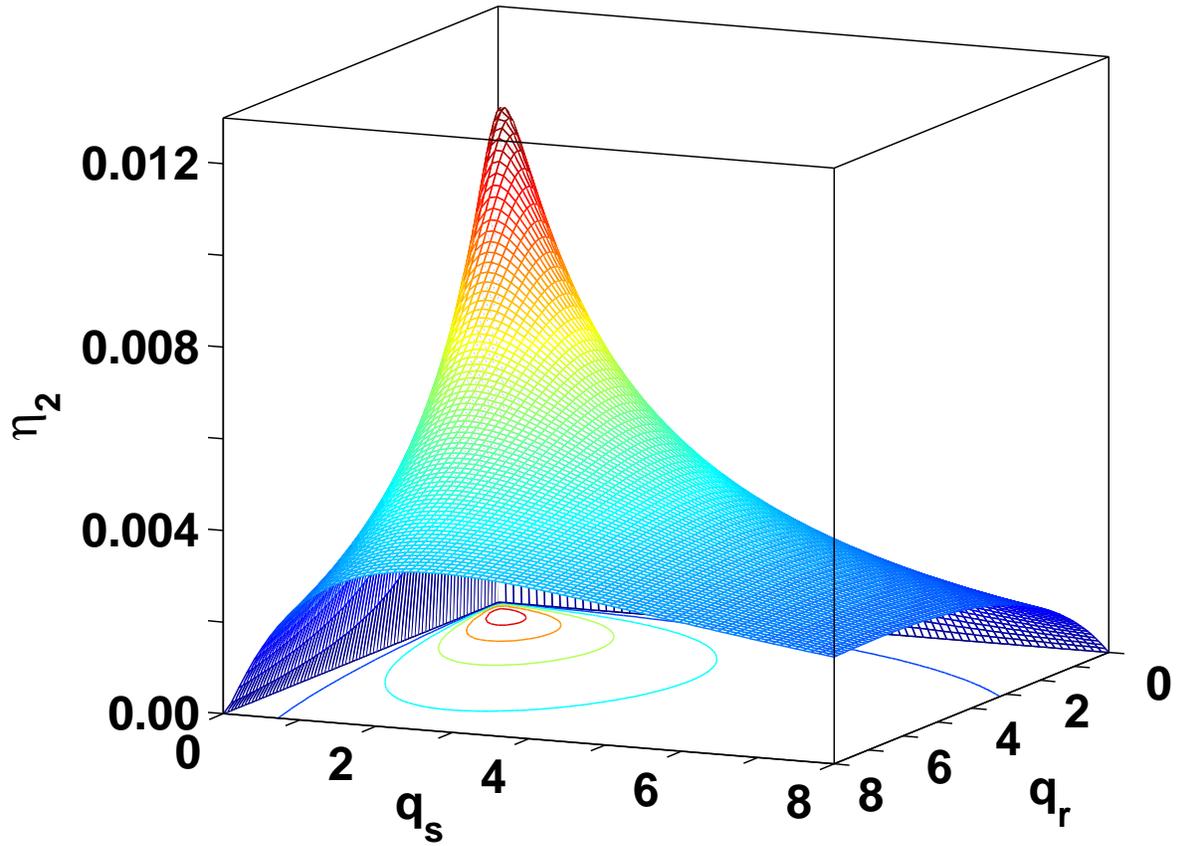}
\caption{3D plot of the efficiency at maximum power relative to the ATP burning 
(given by $g_s\Delta \mu$ with $\Delta \mu=15 k_BT$, see (\ref{effprod})) versus $q_r$ and $q_s$.
The maximum of $\eta_2=1.1\%$ is attained for $(q_r,q_s)=(0.7,0.4)$.} \label{fig6}
\end{figure}

The $q$ factors are defined by
\begin{eqnarray}
q_r=\frac{g_s}{\Gamma_r}=\frac{1}{\Gamma_r\langle t\rangle_r},
\\
q_s=\frac{g_r}{\Gamma_s}=\frac{1}{\Gamma_s\langle t\rangle_s},
\end{eqnarray}
where $\Gamma_r^{-1}$ and $\Gamma_s^{-1}$ are the spring relaxation
times in the R and S states, while $\langle t\rangle_r$ and $\langle t\rangle_s$
are the residence times in R and S. In this subsection we consider the case
where $\langle t\rangle_r << \Gamma_r^{-1}$ and $\langle t\rangle_s << \Gamma_s^{-1}$. 
The motor then switches  rapidly between its internal states and reaches a steady state, where the 
spring has an almost fixed length $L$, which is pulled a little longer during the R-stage and retracts 
by a corresponding amount during the S-stage. 
From (\ref{eq3}) it follows that the equations of motion for the relative coordinate 
have the form
\begin{eqnarray}
&&\dot{y}-\dot{x} = -\Gamma_r(y-x-L_r) +\frac{f}{\zeta_p}~~\text{in the R state},
\\
&&\dot{y}-\dot{x} = -\Gamma_s(y-x-L_s) + \frac{f}{\zeta_v}~~\text{in the S state}.
\end{eqnarray}
The condition that the spring length does not change during a full  R+S cycle is
\begin{eqnarray}
(\Gamma_r(L_r-L)+\frac{f}{\zeta_p})\langle t\rangle_r+ (-\Gamma_s(L-L_s) + \frac{f}{\zeta_v})\langle t\rangle_s=0,\label{Lcond}
\end{eqnarray}
yielding the effective spring length $L$
\begin{eqnarray}
L=\frac{q_sL_r+q_rL_s+cf}{q_r+q_s},
\end{eqnarray}
where
\begin{eqnarray}
c=q_sq_r(\frac{\langle t\rangle_r}{\zeta_p}+\frac{\langle t\rangle_s}{\zeta_v}).
\end{eqnarray} 
The work done during a cycle is given as the force $f$ times the distance travelled by the center of mass. Since 
$x$ and $y$ move in unison we insert the equation of motion for $y$ in (\ref{eq2})
\begin{eqnarray}
(\langle t\rangle_r+\langle t\rangle_s)p & = & f(-\langle t\rangle_r\frac{k_r}{2\zeta_v}(L-L_r)-\langle t\rangle_s\frac{k_s}{2\zeta_v}(L-L_s))\nonumber \\
& = & \frac{f}{2\zeta_v(q_r+q_s)}(a-bf),
\end{eqnarray}
where the parameter $a$ and $b$ are given by
\begin{eqnarray}
a & = & (k_r\langle t\rangle_r q_r - k_s\langle t\rangle_s q_s)(L_r-L_s) = \tilde{\zeta}_v(L_r-L_s)
\\
b & = & c(k_r\langle t\rangle_r + k_s\langle t\rangle_s)=\frac{c\zeta_v}{q_rq_s}(2\tilde{q}_s+q_r).
\end{eqnarray} 
and
\begin{eqnarray}
\tilde{\zeta}_v=\zeta_v\frac{\zeta_p-\zeta_v}{\zeta_p+\zeta_v}, \\
\tilde{q}_s=q_s\frac{\zeta_p}{\zeta_p+\zeta_v}.
\end{eqnarray}
For the maximum power $p_{\text{max}}$ at force $f_{\text{max}}$ we then have
\begin{eqnarray}
&&f_{\text{max}} =  \frac{a}{2b} \label{fmaxAS} 
\\
&&(\langle t\rangle_r+\langle t\rangle_s)p_{\text{max}} = \frac{a^2}{8b(q_r+q_s)}=\frac{\tilde{\zeta}_v^2}{\zeta_v^2}\frac{q_rq_s(L_r-L_s)^2}{8c(q_r+q_s)(2\tilde{q}_s+q_r)} \label{pmaxAS}.
\end{eqnarray}
The dissipated energy is given by
\begin{eqnarray}
 (\langle t\rangle_r+\langle t\rangle_s)q_{\text{out}} &=  
 \langle t\rangle_r\left[\frac{1}{\zeta_p}(\frac{k_r}{2}(L-L_r)-f)^2+\frac{1}{\zeta_v}(\frac{k_r}{2}(L-L_r))^2\right] + \nonumber 
 \\
 &\langle t\rangle_s\frac{1}{\zeta_v}\left[(\frac{k_s}{2}(L-L_s)-f)^2+(\frac{k_s}{2}(L-L_s))^2\right].
\end{eqnarray}
Inserting the relations
\begin{eqnarray}
   &&L-L_r = -(L_r-L_s)q_r\frac{1-\phi_1}{q_r+q_s},
   \\
   &&L-L_r-2f_{\text{max}}/k_r = -(L_r-L_s)q_r\frac{1-\phi_2} {q_r+q_s},
   \\
   &&L-L_s =(L_r-L_s)q_s\frac{1+\phi_3}{q_r+q_s},
   \\
   &&L-L_s-  2f_{\text{max}}/k_s = (L_r-L_s)q_s\frac{1+\phi_4}{q_r+q_s},
\end{eqnarray}
where
\begin{eqnarray}
&&\phi_1 = \frac{\tilde{\zeta}_vq_s}{2\zeta_v(2\tilde{q}_s+q_r)},
\\
&&\phi_2 = \phi_1(1-2\frac{q_r+q_s}{ck_r}),
\\
&&\phi_3 = \frac{\tilde{\zeta}_vq_r}{2\zeta_v(2\tilde{q}_s+q_r)},
\\
&&\phi_4 = \phi_3(1-2\frac{q_r+q_s}{ck_s}),
\end{eqnarray}
we finally arrive at
\begin{eqnarray}
\label{Qapp}
&&(\langle t\rangle_r+\langle t\rangle_s)q_{\text{out}} =
\nonumber 
\\ 
&& \frac{(L_r-L_s)^2}{4(q_r+q_s)^2}\left[\frac{2q_rk_r}
{1+\zeta_p/\zeta_v}\left(\frac{\zeta_v}{\zeta_p}(1-\phi_2)^2+(1-\phi_1)^2\right)+q_sk_s((1-\phi_4)^2+(1-\phi_3)^2)\right].
\end{eqnarray}
It follows from this expression that scaling the residence times by the same factor leads to invariant expressions 
for $p_{\text{max}}$ and $q_{\text{out}}$ which explains the linear contours of constant EMP seen at large $q$'s in Fig.~\ref{fig4}.
In Fig.~\ref{fig5} the approximate EMP ($=1/(1+q_{\text{out}}/p_{\text{max}}$) corresponding to Eqs. (\ref{pmaxAS}) and (\ref{Qapp}) is plotted. The similarity with Fig.~\ref{fig3} is striking.
\acknowledgments We are grateful to A. Imparato for elucidating discussions.
This work has been supported by grants from The Danish Research Council.
\newpage
%
\newpage
\end{document}